\documentclass[useAMS,usenatbib,letterpaper]{mnras}

\usepackage{epsfig}
\usepackage{amsmath}
\usepackage{natbib}
\usepackage{aas_macros,amssymb}
\usepackage{tikz}

\DeclareMathOperator{\sech}{sech}

\begin{document}

\title[Ising Galaxy Bias Model]{\LARGE An Ising model for galaxy bias}

\author[A. Repp \& I. Szapudi]{Andrew Repp\ \& Istv\'an Szapudi\\Institute for Astronomy, University of Hawaii, 2680 Woodlawn Drive, Honolulu, HI 96822, USA}

\date{\today; to be submitted to MNRAS}
\label{firstpage}
\pagerange{\pageref{firstpage}--\pageref{lastpage}}
\maketitle

\begin{abstract}
A reliable model of galaxy bias is necessary for interpreting data from future dense galaxy surveys. Conventional bias models are inaccurate, in that they can yield unphysical results ($\delta_g < -1$) for voids that might contain half of the available cosmological information. For this reason, we present a physically-motivated bias model based on an analogy with the Ising model. With only two free parameters, the model produces sensible results for both high- and low-density regions. We also test the model using a catalog of Millennium Simulation galaxies in cubical survey pixels with side lengths from $2h^{-1}$--$31h^{-1}$Mpc, at redshifts from 0 to 2. We find the Ising model markedly superior to linear and quadratic bias models on scales smaller than $10h^{-1}$Mpc, while those conventional models fare better on scales larger than $30h^{-1}$Mpc. While the largest scale where the Ising model is applicable might vary for a specific galaxy catalog, it should be superior on any scale with a non-negligible fraction of cells devoid of galaxies.
\\
\end{abstract}

\section{Introduction}
\label{sec:intro}
Galaxy surveys represent an important observational constraint on cosmology; indeed, one of the  main science drivers for planned surveys such as \emph{Euclid} \citep{Euclid} and WFIRST \citep{WFIRST} is the expectation that their data will encode large amounts of information on the properties of dark energy. Furthermore, voids -- comprising roughly half the initial volume of the universe -- contain up to half of the cosmological information borne by matter-clustering statistics (see, e.g., \citealp{WCS2015Forecast}). Thus, in order to extract cosmological information from survey data, we require techniques applicable to both clusters and voids.

However, galaxies are biased tracers of matter (see the theoretical treatment in \citealp{Kaiser1984, Bardeen1986}), and thus cosmological inference from galaxy surveys requires modeling of the relationship between the matter and galaxy overdensities ($\delta = \rho/\overline{\rho} - 1$  and $\delta_g = N_\mathrm{gal}/\overline{N} - 1$, respectively). Perhaps the most common approach (e.g., \citealp{Hoffmann2017}) is to expand $\delta_g$ either linearly ($\delta_g = b\delta$) or quadratically ($\delta_g = b_1 \delta + b_2\delta^2/2$). Another approach (e.g., \citealp{delaTorre2013}) performs the expansion in log space so that $\ln(1+\delta_g) = b \ln(1+\delta)$.

One could also attempt to derive a galaxy-dark matter relationship from full-scale halo models and/or hydrodynamical simulations. However, halo modeling is nontrivial, and simulations rely on assumptions about galaxy formation and evolution, many aspects of which remain uncertain. Furthermore, the vast range of scales involved requires simplifications, of unknown impact, in order to render the computations tractable.

Hence, considering the bias models listed above, the first (linear bias) emerges from linear perturbation theory (e.g., \citealp{Desjacques2018}) and thus provides a reasonable description on large scales. However, future surveys (such as \emph{Euclid} and WFIRST) will require models that are accurate on smaller, non-linear scales, and on such scales the standard bias models become problematic. For instance, \citet{Neyrinck2014} note an exponential decline of dark matter haloes at low densities, a decline not captured by a linear bias. In particular, linear and quadratic models easily yield non-physical results ($\delta_g < -1$) in voids, and thus these models fail in regions that potentially constitute half of a survey's information on dark energy. Thus these models fail in regions that potentially constitute half of a survey's information on dark energy. Voids are expected to play an increasingly significant role in cosmological constraints \citep{PisaniEtal2019}. Likewise, in order to detect any screening effects of modified gravity, one must analyze both ends of the density spectrum, again requiring a model of galaxy bias that yields reasonable results at both density extremes.

In this work we present a model (inspired by the Ising model of ferromagnetism) which meets these conditions (Section~\ref{sec:model}). We then analyze its accuracy compared to simulation results (Section~\ref{sec:sims}); discussion and conclusion follow in Sections~\ref{sec:disc} and \ref{sec:concl}.

\section{The Ising Model}
\label{sec:model}

In formulating this model, we focus on dark matter subhaloes, which can host individual galaxies, rather than on the larger haloes (which potentially host many galaxies).

We also make a few simple assumptions about galaxy formation. First, we assume that whether or not a galaxy forms in a particular subhalo depends (to first order) only on initial densities and local physics -- thus ignoring any tidal influences. Since $\ln(1 + \delta)$ captures the approximate initial conditions \citep{NSS09,CarronSzapudi2013}, we express our model in terms of $A\equiv\ln(1+\delta)$.

Second, for the galaxies under consideration in a given survey, we assume it is legitimate to treat the subhaloes in which they form as roughly equivalent. Hence we assume that we can, to first order, characterize these subhaloes as identical entities, each of which is in one of two possible states -- namely, either hosting a galaxy or not.

Third, we note that the release of gravitational potential energy during galaxy formation results in energetic favorability for the ``hosting'' state. And since clustered galaxies collectively occupy deeper potential wells than isolated galaxies, we can extend this assumption to argue that galaxy formation is increasingly favorable in survey cells of higher overall density.

It is now straightforward to map these assumptions on to the Ising model of ferromagnetism, with subhaloes playing the role of atoms. First, an Ising model typically allows interaction between neighboring atoms (facilitating the formation of ferromagnetic domains); however, pursuant to our locality assumption we set the interaction term to zero. Second, an atom in the Ising model has two possible states (spin-up or spin-down); replacing the atoms with subhaloes, the two spin states are analogous to two occupation states (i.e., either hosting a galaxy or not). Third, an Ising model allows external fields to render one of the spin states energetically favorable; analogously, we can treat one of the subhalo occupation states (namely, the galaxy-hosting state) as preferable on energetic grounds.

We thus propose an interactionless Ising model of galaxy incidence in which occupation states replace spin states. These assumptions then yield  (e.g., \citealp{Pathria}) the following expression for the fraction of subhaloes in a favorable (galaxy-hosting) state:
\begin{equation}
f_\mathrm{gal}  =  \frac{1}{2} e^{\beta E} \sech \beta E  =  \frac{1}{1 + e^{-2\beta E}},
\label{eq:Ising_no_J}
\end{equation}
where $E$ is energy difference between the two states and $\beta$ is analogous to inverse temperature. (The thermodynamics of gravitational collapse suggest that this temperature analogue will be negative.) The unitless quantity $\beta E$ depends (pursuant to our first and third assumptions) on the initial density, for which we use the log density $A \equiv \ln(1+\delta)$ as a proxy \citep{CarronSzapudi2013}. We thus substitute a linear function of $A$ for $\beta E$ and find that Equation~\ref{eq:Ising_no_J} reduces to a Fermi-Dirac distribution:
\begin{equation}
f_\mathrm{gal} = \frac{1}{1 + \exp\left(\frac{A-A_t}{-T}\right)}\hspace{1cm}(T>0).
\label{eq:FD}
\end{equation}
In this expression, $A_t$ marks the transition density between empty and occupied subhaloes. The subhaloes in higher-density cells fill up first, and $T$ (the absolute value of the negative ``temperature'') parametrizes the sharpness of the transition from the empty to the occupied state: as $T$ approaches zero, Equation~\ref{eq:FD} approaches a step function.

Note that the Fermi-Dirac form of Equation~\ref{eq:FD} suggests other analogies: one could argue that galaxies observe an exclusion principle, with at most one galaxy occupying a given subhalo. One could also treat the transition density $A_t$ as analogous to chemical potential since it characterizes the (log) density at which the next galaxy would form.

Continuing, our goal is to describe the expected number of galaxies per survey cell as a function of the underlying dark matter density. We would expect (\emph{ceteris paribus}) the number of subhaloes in a survey cell to be proportional to the cell's matter density ($\langle N_\mathrm{sh} \rangle_A \propto 1 + \delta \equiv e^A$). Hence, given a cell of log density $A$, we express the occupied fraction $f_\mathrm{gal}$ of subhaloes as follows:
\begin{equation}
f_\mathrm{gal} = \frac{\langle N_\mathrm{gal} \rangle_A}{\langle N_\mathrm{sh}\rangle_A} = \frac{\langle N_\mathrm{gal} \rangle_A}{b\overline{N}(1 + \delta)},
\end{equation}
where $\overline{N}$ is the global mean number of galaxies per cell. (Here, as in the rest of this paper, subscripted $A$ indicates the underlying log dark matter density; thus $\langle N_\mathrm{gal} \rangle_A$ denotes the expected number of galaxies in a cell with log density $A$.) Thus it is convenient to write Equation~\ref{eq:FD} in terms of $M$, the expected number of galaxies per dark matter mass:
\begin{equation}
\label{eq:final_FD}
 M \equiv \langle N_\mathrm{gal} \rangle_A \cdot (1 + \delta)^{-1} = \frac{b\overline{N}}{1 + \exp\left(\frac{A_t-A}{T}\right)}\hspace{1cm}(T>0),
 \end{equation}
where, again, $A \equiv \ln(1+\delta)$.
In high-density regions $A \gg A_t$, so that $M$ approaches $b\overline{N}$, and $\langle N_\mathrm{gal} \rangle_A$ approaches $b\overline{N}(1+\delta)$; thus at high densities the number of galaxies is proportional to the amount of underlying matter. For low-density regions ($A \ll A_t$), the number of galaxies drops exponentially to zero, as \citet{Neyrinck2014} observe.

Note that the parameter $b$ represents the overall bias in high-density regions; in these regions, $(1+\delta_g) = \langle N_\mathrm{gal} \rangle_A/ \overline{N} = b(1+\delta)$, so that $\delta_g \sim b \delta$, as in the linear bias model. Since high-density regions are the predominant influence on the matter power spectrum, it is not surprising that linear bias models seem to fit the relationship between matter and galaxy spectra. However, both of these spectra -- and the linear bias model, as noted above -- discard much of the (substantial) information inherent in voids \citep{NSS09,CarronSzapudi2013,CarronSzapudi2014,WCS2015,Repp2015}.

Finally, we can derive an additional constraint -- and thus reduce the number of free model parameters to two -- by considering the (global) mean number of galaxies. We obtain this mean by integrating the expected galaxy counts from Equation~\ref{eq:final_FD} against the probability distribution $\mathcal{P}(A)$ of the underlying matter density:
\begin{equation}
\overline{N} = \int dA\, \mathcal{P}(A) \langle N_\mathrm{gal} \rangle_A = \int dA\, \mathcal{P}(A) \frac{b\overline{N}e^A}{1 + \exp\left(\frac{A_t-A}{T}\right)}.
\end{equation}
It follows that
\begin{equation}
b = \left( \int dA\,\, \mathcal{P}(A) \frac{e^A}{1 + \exp\left(\frac{A_t-A}{T}\right)} \right)^{-1}.
\label{eq:constraint}
\end{equation}
To impose this constraint we must know the dark matter probability distribution. Since in this work we consider simulation results -- in which we have access to both the dark matter and the galaxy contents of each cell -- we shall for this paper use the empirical dark matter distribution given by the simulation itself.

Before proceding to comparison with simulation results, we state explicitly the following analogues of Equation~\ref{eq:final_FD} for the linear, quadratic, and logarithmic bias models, where again we define $M\equiv \langle N_\mathrm{gal} \rangle_A \cdot (1+\delta)^{-1}$:
\begin{align}
M & = \overline{N} \left(b + (1-b)e^{-A} \right) \hspace{5mm}\mbox{(linear bias)}\label{eq:lin}\\
M & = \overline{N} \left( \frac{b_2}{2}e^A + \left(b_1-b_2\right) + e^{-A}\left( \frac{b_2}{2} - b_1 + 1 \right)\right)\\
  &   \hspace{5cm} \mbox{    (quadratic bias)}\nonumber\\
M & = \overline{N} e^{A(b-1)}\hspace{5mm}\mbox{(logarithmic bias)}\label{eq:log}
\end{align}
We fit Equations~\ref{eq:final_FD} and \ref{eq:lin}--\ref{eq:log} to simulation results in the following section.

\section{Comparison to Simulation Results}
\label{sec:sims}
To validate the Ising model of Equation~\ref{eq:final_FD}, we must compare it to simulations and/or observations. This section accomplishes the former by reporting a series of comparisons to the Millennium Simulation \citep{Springel2005}. In a subsequent paper (Repp \& Szapudi, in prep.) we perform comparisons to observational data.

The Millennium Simulation\footnote{http://gavo.mpa-garching.mpg.de/Millennium/} provides dark matter densities in a cubical volume with sides of $500h^{-1}$ Mpc. We consider simulation snapshots 32, 41, 48, and 63, corresponding in the original Millennium Simulation cosmology to $z=2.07$, 0.99, 0.51, and 0.00, respectively.

For our galaxy catalog, we employ the results described in \citet{Bertone2007}, in which the L-Galaxies semi-analytic model (also known as the Munich model, outlined in \citealt{Croton2006} and \citealt{DeLucia2006}) is applied to the Millennium Simulation outputs. To obtain a dense sample, we impose the stellar mass cut $M_\star \ge 10^9h^{-1}M_\odot$. We then determine counts-in-cells (for both the dark matter and galaxy distributions) using cubical cells with side lengths ranging from $1.95h^{-1}$ Mpc to $31.25h^{-1}$ Mpc; in this way we obtain results for five different smoothing scales. (The smallest scale yields $256^3$ cells; the largest scale, $16^3$.) In this manner, we construct galaxy catalogs for each of the above four redshifts. At $z=0$, the mean number $\overline{N}$ of galaxies per cell ranges from 0.352 on the smallest scale to almost 1450 on the largest; at $z=2.1$, $\overline{N}$ ranges from 0.275 to approximately 1120.

\subsection{Fits to Simulation Data}
\label{sec:dircomp}
\begin{figure*}
\leavevmode\epsfxsize=16cm\epsfbox{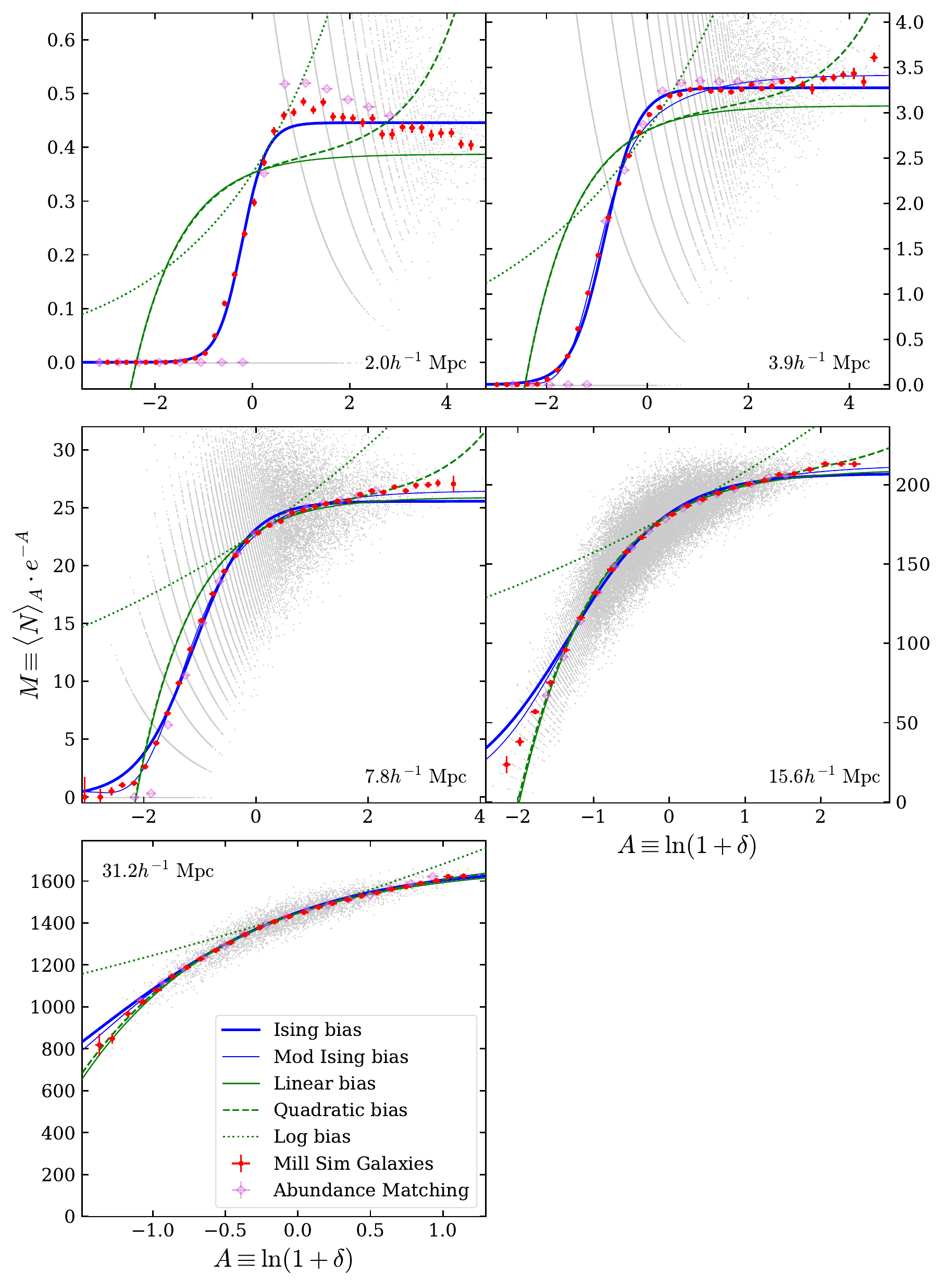}
\caption{Fits of various galaxy bias models to Millennium Simulation results at $z=0$: light grey points show the galaxy counts $N_\mathrm{gal}$ and the log dark matter density $A=\ln(1+\delta)$ on five smoothing scales (i.e., side-lengths of cubical pixels). Note that we normalize the vertical axis by the dark matter density $e^A$, so that the vertical axis represents the average number of galaxies per dark matter mass ($M$ in Equation~\ref{eq:final_FD}). The red data points show $\langle N_\mathrm{gal} \rangle_A \cdot e^{-A}$ in each bin of $A$-values. The thick blue curves show the best-fitting Ising bias model at each scale, and the green curves show the best-fitting linear, quadratic, and logarithmic bias models at each scale. The thin blue curves show the best fit for the modified Ising model of Section~\ref{sec:modIsing}. Light magenta points show the results of the abundance matching procedure of Section~\ref{sec:AM}.}
\label{fig:MillSim_scat0}
\end{figure*}

\begin{figure*}
\leavevmode\epsfxsize=16cm\epsfbox{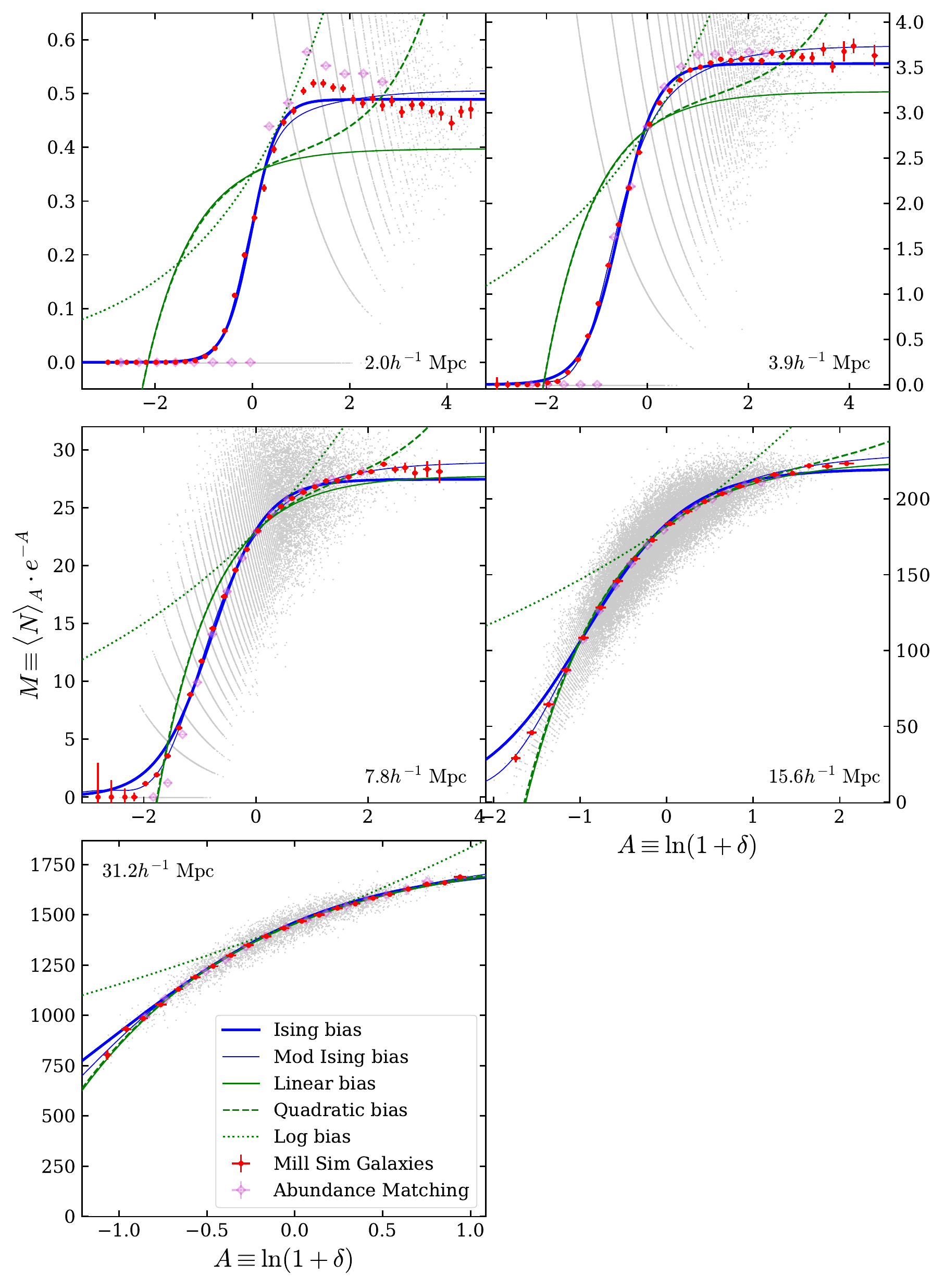}
\caption{Fits of various galaxy bias models to Millennium Simulation results at $z=0.51$; see caption of Figure~\ref{fig:MillSim_scat0} for a full description.}
\label{fig:MillSim_scat0.5}
\end{figure*}

\begin{figure*}
\leavevmode\epsfxsize=16cm\epsfbox{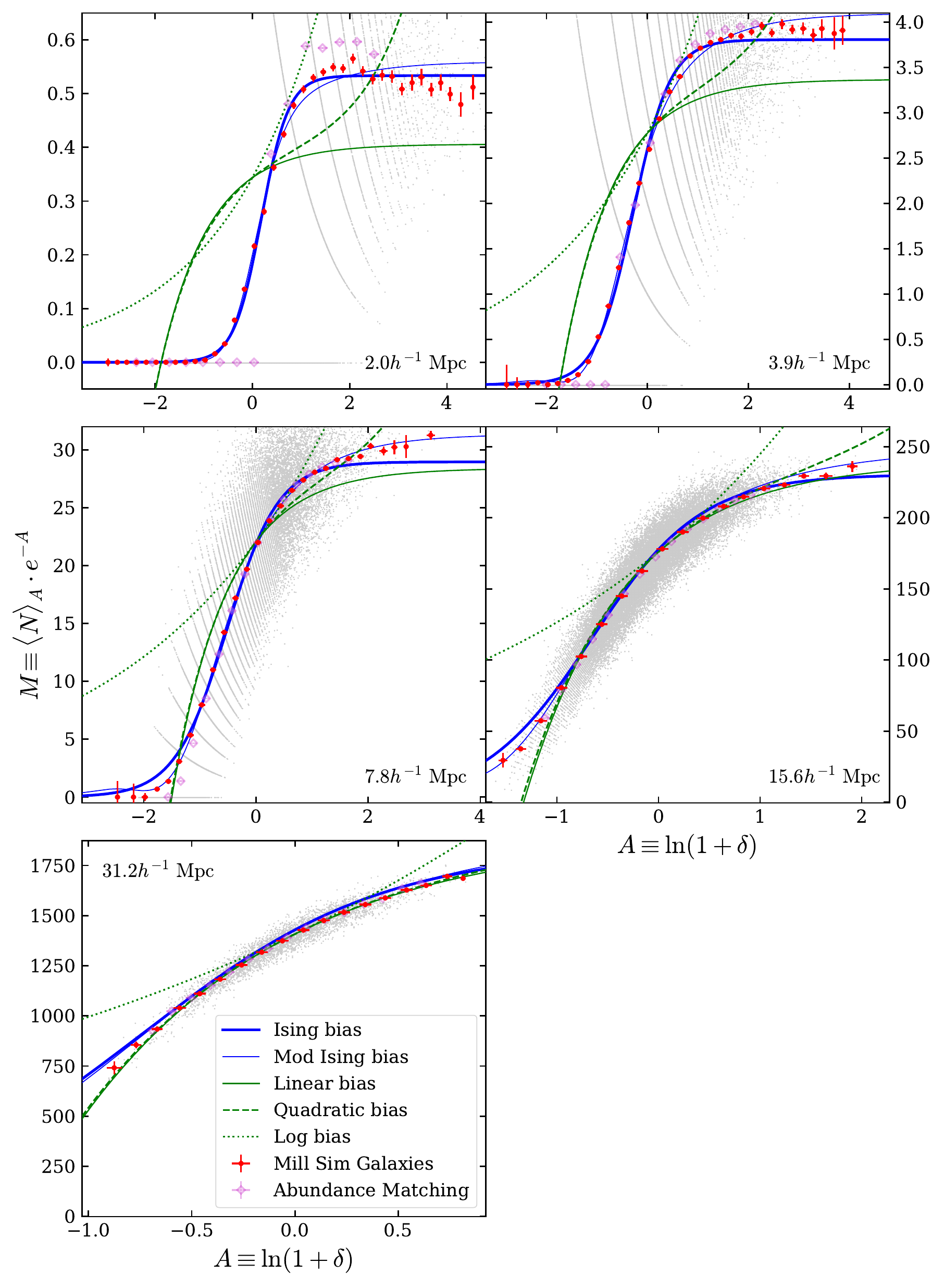}
\caption{Fits of various galaxy bias models to Millennium Simulation results at $z=0.99$; see caption of Figure~\ref{fig:MillSim_scat0} for a full description.}
\label{fig:MillSim_scat1.0}
\end{figure*}

\begin{figure*}
\leavevmode\epsfxsize=16cm\epsfbox{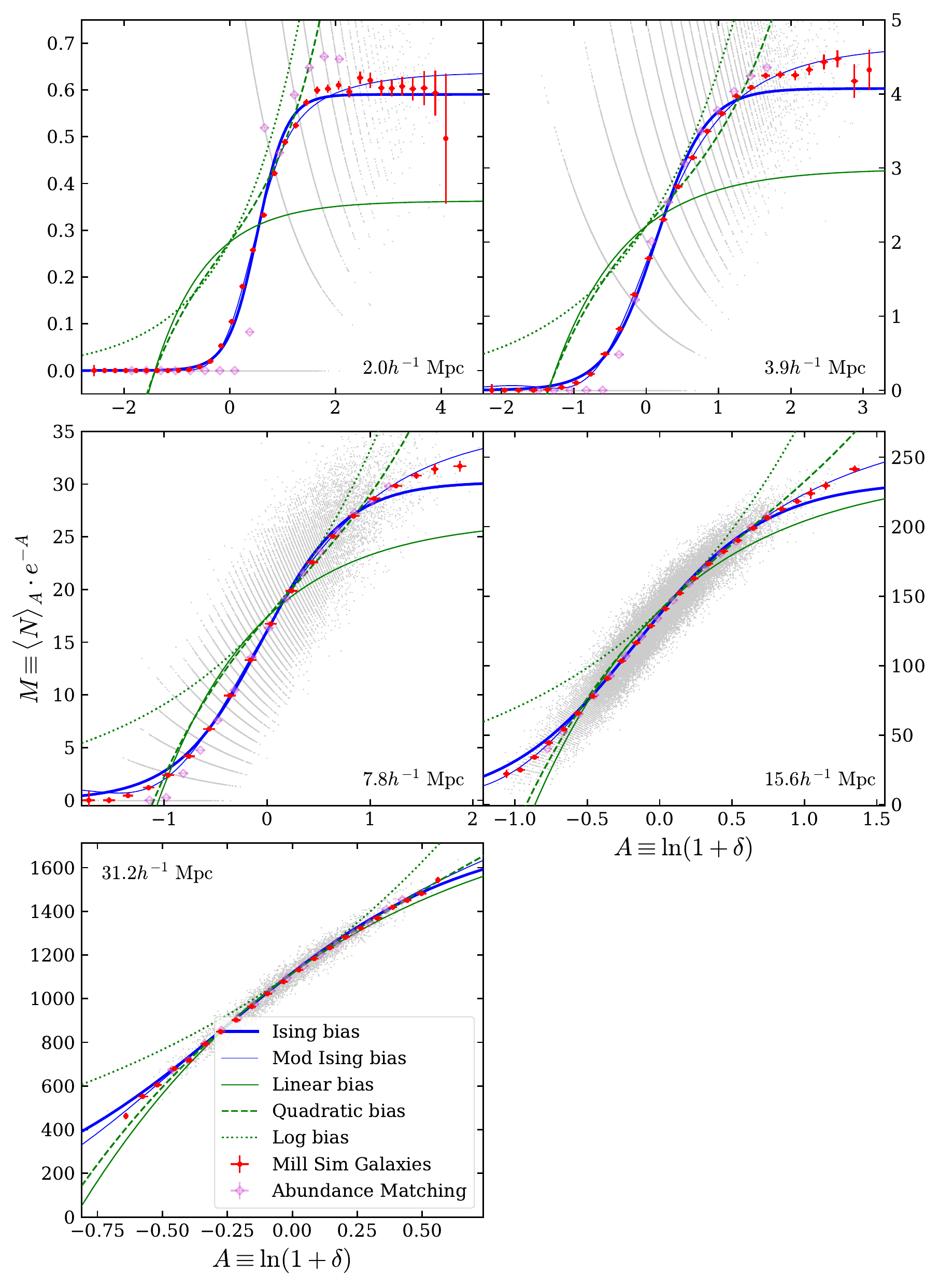}
\caption{Fits of various galaxy bias models to Millennium Simulation results at $z=2.07$; see caption of Figure~\ref{fig:MillSim_scat0} for a full description.}
\label{fig:MillSim_scat2.1}
\end{figure*}

Figure~\ref{fig:MillSim_scat0} displays in light gray (for each of the five cell sizes) the counts-in-cells values for $A$ and $N_\mathrm{gal}$ at $z=0$; Figures~\ref{fig:MillSim_scat0.5}--\ref{fig:MillSim_scat2.1} do the same for redshifts 0.5, 1.0, and 2.1. Note that, as in Equation~\ref{eq:final_FD}, we normalize the galaxy counts by $e^A \equiv 1+\delta$. In the first three panels of the figures, the loci corresponding to $N_\mathrm{gal}=0,1,\ldots$ are evident, with the curvature of the $N_\mathrm{gal} \ge 1$ loci being the result of the aforementioned normalization. These plots show clearly the large degree of stochasticity on scales smaller than $\sim 10h^{-1}$Mpc, in that for a given dark matter density $A$ there is a wide range of resulting galaxy counts $N_\mathrm{gal}$. Thus our first two assumptions in Section~\ref{sec:model} are true only in an average sense -- a nuance reflected in Equation~\ref{eq:final_FD} by the use of $\langle N_\mathrm{gal} \rangle_A$, the mean number of galaxies for a given matter density.

\begin{figure}
\leavevmode\epsfxsize=8cm\epsfbox{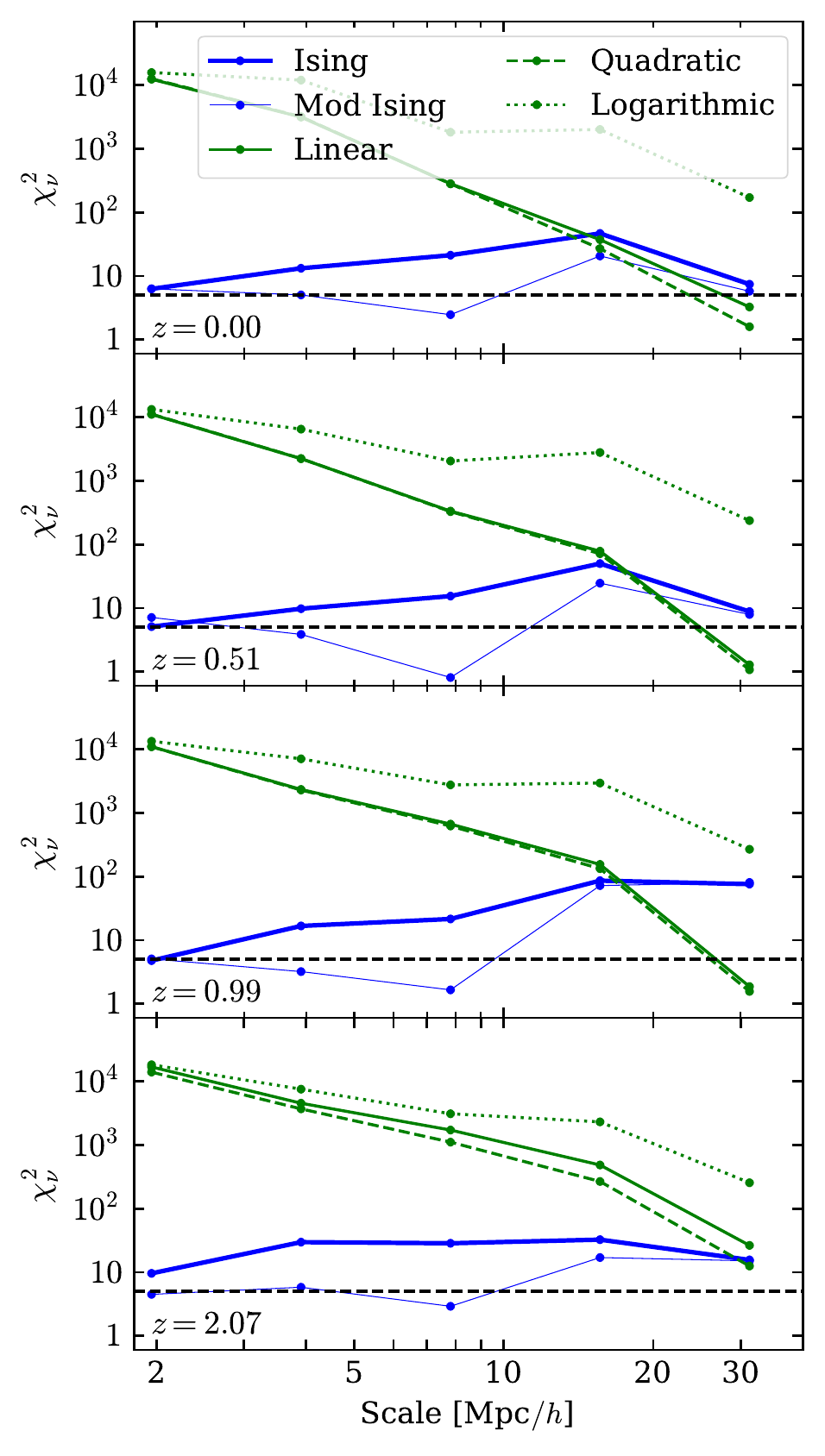}
\caption{Best-fitting $\chi^2$ per degree of freedom for various models at various redshifts. The thin blue lines show the results for the modified Ising model of Section~\ref{sec:modIsing}. The horizontal dashed lines show (for reference) the location of $\chi^2_\nu = 5$.}
\label{fig:chi2dof}
\end{figure}

\begin{table*}
\begin{tabular}{cccccccc}
\hline\\[-6mm]
Scale & Bias  &                                          &                        &            & Bias     &  Best-fitting                  &\rule{0cm}{15pt} \\
(Mpc$/h$) & Model & Best-fitting Parameters                      & $\chi^2_\nu$           &\rule{5mm}{0cm}& Model &  Parameters                & $\chi^2_\nu$\\ \hline\\[-4mm]

$z = 0$:\hspace{1cm} & &                                      &                        &            &          &                            & \\
1.95  & Ising     & $b = 1.27$, $A_t = -0.213$, $T = 0.251$  & $6.26$                 &            & Quadratic& $b_1 = 1.12$, $b_2 = 0.024$ & $1.26 \times 10^4$     \\
      & Linear    & $b = 1.10$                               & $1.24 \times 10^4$     &            & Log      & $b = 1.39$                 & $1.58 \times 10^5$\\
3.91  & Ising     & $b = 1.17$, $A_t = -0.843$, $T = 0.327$  & $13.2$ \rule{0cm}{12pt}&            & Quadratic& $b_1 = 1.10$, $b_2 = 0.009$& $3190$\\
      & Linear    & $b = 1.10$                               & $3150$                 &            & Log      & $b = 1.29$                 & $1.20 \times 10^4$\\
7.81  & Ising     & $b = 1.12$, $A_t = -1.13$, $T = 0.498$   & $21.2$\rule{0cm}{12pt} &            & Quadratic& $b_1 = 1.14$, $b_2 = 0.009$& $281$\\
      & Linear    & $b = 1.13$                               & $282$                  &            & Log      & $b = 1.14$                 & $1820$\\
15.6  & Ising     & $b = 1.15$, $A_t = -1.33$, $T = 0.671$   & $46.6$\rule{0cm}{12pt} &            & Quadratic& $b_1 = 1.16$, $b_2 = 0.010$& $26.9$\\
      & Linear    & $b = 1.16$                               & $37.2$                 &            & Log      & $b = 1.14$                 & $2010$\\
31.3  & Ising     & $b = 1.16$, $A_t = -1.48$, $T = 0.788$   & $7.40$\rule{0cm}{12pt} &            & Quadratic& $b_1 = 1.16$, $b_2 = 0.016$& $1.59$\\
      & Linear    & $b = 1.16$                               & $3.25$                 &            & Log      & $b = 1.15$                 & $171$\\
      
$z = 0.51$:\hspace{1cm} & &                                  &                        &            &          &                            & \\
1.95  & Ising     & $b = 1.39$, $A_t = -0.019$, $T = 0.268$  & $5.10$                 &            & Quadratic& $b_1 = 1.15$, $b_2 = 0.043$ & $1.12 \times 10^4$     \\
      & Linear    & $b = 1.13$                               & $1.11 \times 10^4$     &            & Log      & $b = 1.42$                 & $1.33 \times 10^4$\\
3.91  & Ising     & $b = 1.26$, $A_t = -0.537$, $T = 0.357$  & $9.76$ \rule{0cm}{12pt}&            & Quadratic& $b_1 = 1.15$, $b_2 = 0.018$& $2240$\\
      & Linear    & $b = 1.15$                               & $2230$                 &            & Log      & $b = 1.30$                 & $6500$\\
7.81  & Ising     & $b = 1.20$, $A_t = -0.800$, $T = 0.482$  & $15.4$\rule{0cm}{12pt} &            & Quadratic& $b_1 = 1.22$, $b_2 = 0.019$& $328$\\
      & Linear    & $b = 1.21$                               & $334$                  &            & Log      & $b = 1.21$                 & $2050$\\
15.6  & Ising     & $b = 1.21$, $A_t = -0.955$, $T = 0.590$  & $50.0$\rule{0cm}{12pt} &            & Quadratic& $b_1 = 1.25$, $b_2 = 0.015$& $71.5$\\
      & Linear    & $b = 1.24$                               & $78.4$                 &            & Log      & $b = 1.21$                 & $2790$\\
31.3  & Ising     & $b = 1.20$, $A_t = -1.06$, $T = 0.650$   & $8.80$\rule{0cm}{12pt} &            & Quadratic& $b_1 = 1.24$, $b_2 = 0.010$& $1.07$\\
      & Linear    & $b = 1.24$                               & $1.28$                 &            & Log      & $b = 1.231$                & $237$\\
            
$z = 0.99$:\hspace{1cm} & &                                  &                        &            &          &                            & \\
1.95  & Ising     & $b = 1.55$, $A_t = 0.183$, $T = 0.280$   & $4.76$                 &            & Quadratic& $b_1 = 1.21$, $b_2 = 0.071$& $1.10 \times 10^4$     \\
      & Linear    & $b = 1.18$                               & $1.10 \times 10^4$     &            & Log      & $b = 1.48$                 & $1.34 \times 10^4$\\
3.91  & Ising     & $b = 1.37$, $A_t = -0.281$, $T = 0.366$  & $16.7$ \rule{0cm}{12pt}&            & Quadratic& $b_1 = 1.23$, $b_2 = 0.050$& $2290$\\
      & Linear    & $b = 1.21$                               & $2340$                 &            & Log      & $b = 1.38$                 & $7090$\\
7.81  & Ising     & $b = 1.31$, $A_t = -0.520$, $T = 0.462$  & $21.5$\rule{0cm}{12pt} &            & Quadratic& $b_1 = 1.30$, $b_2 = 0.047$& $623$\\
      & Linear    & $b = 1.29$                               & $667$                  &            & Log      & $b = 1.30$                 & $2760$\\
15.6  & Ising     & $b = 1.31$, $A_t = -0.657$, $T = 0.540$  & $86.1$\rule{0cm}{12pt} &            & Quadratic& $b_1 = 1.36$, $b_2 = 0.043$& $133$\\
      & Linear    & $b = 1.36$                               & $155$                  &            & Log      & $b = 1.33$                 & $2950$\\
31.3  & Ising     & $b = 1.30$, $A_t = -0.733$, $T = 0.579$  & $75.9$\rule{0cm}{12pt} &            & Quadratic& $b_1 = 1.36$, $b_2 = 0.018$& $1.56$\\
      & Linear    & $b = 1.36$                               & $1.86$                 &            & Log      & $b = 1.347$                & $268$\\
      
$z = 2.07$:\hspace{1cm} & &                                  &                        &            &          &                            & \\
1.95  & Ising     & $b = 2.15$, $A_t = 0.525$, $T = 0.277$   & $9.59$                 &            & Quadratic& $b_1 = 1.58$, $b_2 = 0.679$& $1.40 \times 10^4$     \\
      & Linear    & $b = 1.32$                               & $1.69 \times 10^4$     &            & Log      & $b = 1.76$                 & $1.83 \times 10^4$\\
3.91  & Ising     & $b = 1.84$, $A_t = 0.139$, $T = 0.344$   & $29.7$ \rule{0cm}{12pt}&            & Quadratic& $b_1 = 1.51$, $b_2 = 0.436$& $3700$\\
      & Linear    & $b = 1.35$                               & $4540$                 &            & Log      & $b = 1.67$                 & $7500$\\
7.81  & Ising     & $b = 1.74$, $A_t = -0.054$, $T = 0.411$  & $28.5$\rule{0cm}{12pt} &            & Quadratic& $b_1 = 1.66$, $b_2 = 0.470$& $1110$\\
      & Linear    & $b = 1.54$                               & $1720$                 &            & Log      & $b = 1.65$                 & $3100$\\
15.6  & Ising     & $b = 1.67$, $A_t = -0.161$, $T = 0.451$  & $32.5$\rule{0cm}{12pt} &            & Quadratic& $b_1 = 1.76$, $b_2 = 0.337$& $267$\\
      & Linear    & $b = 1.73$                               & $485$                  &            & Log      & $b = 1.70$                 & $2310$\\
31.3  & Ising     & $b = 1.59$, $A_t = -0.239$, $T = 0.452$  & $15.6$\rule{0cm}{12pt} &            & Quadratic& $b_1 = 1.77$, $b_2 = 0.275$& $12.5$\\
      & Linear    & $b = 1.76$                               & $26.4$                 &            & Log      & $b = 1.75$                 & $255$\\
\hline
\end{tabular}
\caption{Results of fitting four bias models (Ising, linear, quadratic, and logarithmic) to the average number of galaxies per dark matter mass ($M = \langle N_\mathrm{gal} \rangle_A \cdot e^{-A}$) as a function of log dark matter density ($A = \ln(1+\delta)$); we derive the dark matter densities and galaxy counts from the Millennium Simulation data and galaxy catalogs. The table shows the best-fitting values (and the corresponding reduced $\chi^2$-values) for the models at various smoothing scales and redshifts; see the corresponding graphical representations in Figures~\ref{fig:MillSim_scat0}--\ref{fig:MillSim_scat2.1}. \label{tab:MS_dat}}
\end{table*}

We proceed to consider these mean values by calculating $\langle N_\mathrm{gal} \rangle_A$ in a set of bins in $A$. To achieve sufficient resolution for estimating the probability distribution $\mathcal{P}(A)$ (without unduly manipulating the bin placement), we begin with a bin width $\Delta A = 0.2$ and then reduce the bin size if necessary to ensure that at least twenty bins contain survey cells. The resulting bin widths are $\Delta A = 0.2$ for all panels of Figures~\ref{fig:MillSim_scat0}--\ref{fig:MillSim_scat2.1}, with the following exceptions: for $31.2h^{-1}$-Mpc cells at $z = 0$, 0.5, and 1.0, we have $\Delta A = 0.1$; for $15.6h^{-1}$-Mpc cells at $z=2.1$ we have $\Delta A = 0.1$; and for $31.2h^{-1}$-Mpc cells at $z=2.1$ we have $\Delta A = 0.06$.

In each bin we then calculate the values of $\langle M \rangle_\mathrm{bin} = \langle N_\mathrm{gal} \cdot e^{-A} \rangle_\mathrm{bin}$ from Equation~\ref{eq:final_FD}; we plot these values in red on Figures~\ref{fig:MillSim_scat0}--\ref{fig:MillSim_scat2.1}. The horizontal error bars for these points depict the standard deviation of the $A$-values in each bin, and the vertical error bars depict the estimated uncertainties of the measured $M$-values. (See Appendix~\ref{app:err} for details, where we explain our use of a subsample to reduce the effect of covariance on the error estimates.) It is clear from the plots that a sigmoid function, such as that provided by the Ising model, is a reasonable approximation on all five scales.

To quantify the fit of the various bias models, we determine for each one the best-fitting parameter values and the corresponding values of $\chi^2_\nu$ (chi-squared per degree of freedom), for each scale and redshift. To impose the net-galaxy constraint (Equation~\ref{eq:constraint}) on the Ising model, we use the empirical probability distribution $\mathcal{P}(A)$ derived from counting cells in the $A$-bins described above. The resulting best-fitting Ising bias models appear as the thick blue curves on Figures~\ref{fig:MillSim_scat0}--\ref{fig:MillSim_scat2.1}, and the best-fitting linear, quadratic, and logarithmic bias models appear in green. The best-fitting parameter values and their corresponding $\chi^2_\nu$-values appear in Table~\ref{tab:MS_dat}, and Figure~\ref{fig:chi2dof} displays the $\chi^2_\nu$-values for the various models.

Considering the reduced chi-squared values, we conclude that though the Ising model is not a perfect fit to the data, its fit is superior to the others -- in many cases, vastly superior -- at scales smaller than $\sim 10h^{-1}$ Mpc, specifically because it avoids unphysical predictions for low densities. In particular, at scales $\la 5h^{-1}$ the linear and quadratic models significantly underestimate the galaxy bias in the high-density regime. At intermediate scales (around $15h^{-1}$ Mpc), none of the models seems to provide a particularly good fit to the simulation results. However, at the largest scale analyzed (around $30h^{-1}$ Mpc) the linear and quadratic models fit the data better than the Ising model.

\subsection{A Modified Ising model}
\label{sec:modIsing}
Figures~\ref{fig:MillSim_scat0}--\ref{fig:MillSim_scat2.1} (especially the panels displaying the 3.9--15.6$h^{-1}$-Mpc scales) seem to show that the Ising model parameters which give the best fit at moderate densities do not necessarily reflect the shape of the data at high densities; indeed, at high densities the linear bias seems to best reflect the curvature of the data points (albeit with a significant horizontal offset). Given this pattern, it is worthwhile to explore, in a preliminary fashion, the utility of modifiying the Ising model to provide better asymptotic behavior at high densities.

Our strategy is to expand the models about $e^{-A}=0$. We note that the linear bias model (Equation~\ref{eq:lin}) is (naturally) strictly first-order in this expansion:
\begin{equation}
M = \overline{N}b + \overline{N}(1-b)e^{-A}.
\label{eq:lin_expand}
\end{equation}
Expanding the Ising model (Equation~\ref{eq:final_FD}) in the same way, we obtain
\begin{equation}
M = \overline{N}b + O(e^{-2A}),
\end{equation}
thus converging in the limit $A \rightarrow \infty$ to the same constant value as Equation~\ref{eq:lin_expand}, but lacking the correct first-order approach to that limit.

We thus desire a simple function $f(e^{-A})$ which approaches $\overline{N}(1-b)e^{-A}$ in the limit $A \rightarrow \infty$ while approaching zero in the limit $A \rightarrow -\infty$. A double exponential yields the correct behavior, so we let
\begin{align}
f(e^{-A}) & = \overline{N}(1-b)e^{-A}\exp\left(-k e^{-A}\right)\\
          & = \overline{N}(1-b)e^{-A} + O(e^{-2A})
\end{align}
serve as our correction term, where $k$ is a free parameter controlling the location of the transition between the high- and low-density regimes. Hence, we can now investigate whether it is advantageous to consider a modified Ising model
\begin{equation}
M = \frac{b\overline{N}}{1 + \exp\left(\frac{A_t-A}{T}\right)} + (1-b)\overline{N}e^{-A}\exp\left(-k e^{-A}\right).
\label{eq:mod_Ising}
\end{equation}
The net-galaxy constraint analogous to Equation~\ref{eq:constraint} now becomes
\begin{equation}
b = \frac{1-I_2}{I_1 - I_2},
\label{eq:modconstraint}
\end{equation}
where $I_1$ and $I_2$ are the integrals
\begin{align}
I_1 & = \int dA\, \mathcal{P}(A) \frac{e^A}{1+\exp\left(\frac{A_t - A}{T}\right)}\\
I_2 & = \int dA\, \mathcal{P}(A) \exp\left(-k e^{-A} \right).
\end{align}
In the limit of $k \rightarrow \infty$, we have $I_2 = 0$, and Equations~\ref{eq:mod_Ising} and \ref{eq:modconstraint} reduce to Equations~\ref{eq:final_FD} and \ref{eq:constraint}, respectively.

We thus repeat the fitting procedure of Section~\ref{sec:dircomp} to investigate the improvement (if any) achieved by the addition of the extra term. The resulting best fits appear as thin blue lines in Figures~\ref{fig:MillSim_scat0}--\ref{fig:MillSim_scat2.1}, and the best-fitting parameters for the modified model appear, along with their reduced $\chi^2$-values, in Table~\ref{tab:mod_Ising}. The $\chi^2_\nu$ values for this model also appear in Figure~\ref{fig:chi2dof} as points connected by thin blue lines.
\begin{table}
\begin{tabular}{cccccc}
\hline\\[-6mm]
  Scale   & \multicolumn{4}{c}{\dotfill Best-fitting Parameters\dotfill}  & \rule{0cm}{15pt} \\
(Mpc$/h$) & $b$ & $A_t$ & $T$ & $k$                                   & $\chi^2_\nu$     \\ \hline\\[-4mm]
$z = 0$:\hspace{1cm} & & & & &\\
1.95  & 1.27 & $-0.211$ & 0.252 & 19.2  & 6.27\\
3.91  & 1.22 & $-1.10$  & 0.352 & 0.416 & 5.03\\
7.81  & 1.16 & $-1.55$  & 0.470 & 0.230 & 2.46\\
15.6  & 1.18 & $-1.31$  & 0.572 & 1.00  & 20.6\\
31.3  & 1.19 & $-1.38$  & 0.703 & 1.51  & 5.75\\

$z = 0.51$:\hspace{1cm}\rule{0cm}{12pt} & & \\
1.95  & 1.44 & $-0.125$ & 0.284 & 1.26  & 7.11\\
3.91  & 1.33 & $-0.800$ & 0.381 & 0.542 & 3.85\\
7.81  & 1.26 & $-1.22$  & 0.468 & 0.339 & 0.808\\
15.6  & 1.27 & $-1.45$  & 0.494 & 0.297 & 24.6\\
31.3  & 1.25 & $-1.67$  & 0.509 & 0.217 & 7.92\\

$z = 0.99$:\hspace{1cm}\rule{0cm}{12pt} & & \\
1.95  & 1.63 & $0.049$  & 0.318 & 1.32  & 5.07\\
3.91  & 1.48 & $-0.593$ & 0.426 & 0.610 & 3.20\\
7.81  & 1.42 & $-0.976$ & 0.485 & 0.431 & 1.65\\
15.6  & 1.41 & $-1.09$  & 0.492 & 0.442 & 72.0\\
31.3  & 1.39 & $-0.874$ & 0.492 & 0.816 & 80.7\\

$z = 2.07$:\hspace{1cm}\rule{0cm}{12pt} & & \\
1.95  & 2.32 & $0.388$  & 0.344 & 1.69  & 4.48\\
3.91  & 2.11 & $-0.154$ & 0.461 & 0.897 & 5.80\\
7.81  & 2.06 & $-0.489$ & 0.532 & 0.677 & 2.91\\
15.6  & 1.98 & $-0.643$ & 0.548 & 0.615 & 17.0\\
31.3  & 1.93 & $-0.822$ & 0.563 & 0.500 & 15.1\\
\hline
\end{tabular}
\caption{Results of fitting a modified Ising model (Equation~\ref{eq:mod_Ising}) to dark matter densities and galaxy counts from the Millennium Simulation data and galaxy catalogs. The table shows the best-fitting values (and the corresponding reduced $\chi^2$-values) for the modified model at various smoothing scales and redshifts; see the corresponding graphical representations (thin blue curves) in Figures~\ref{fig:MillSim_scat0}--\ref{fig:chi2dof}. \label{tab:mod_Ising}}
\end{table}

The utility of including the modification term (with its extra free parameter $k$) seems to be mixed. At the smallest scale (around $2h^{-1}$ Mpc) and the largest scale (around $30h^{-1}$ Mpc), it provides essentially no improvement; in particular, at $30h^{-1}$-Mpc scales the quadratic model still consistently outperforms the Ising model. However, at intermediate scales (roughly $5$--$10h^{-1}$ Mpc) there is significant improvement, with $\chi^2_\nu$ values in many cases comparable to unity. At scales around $15h^{-1}$ Mpc, the modification does improve the fit -- but as before, none of the models seems to yield a particularly good description of the simulation results.

Thus there seems to a non-negligible scale-dependence in the shape of the matter-galaxy relationship. This phenomenon merits further investigation which we, however, leave to future work. In addition, unlike the ``plain'' Ising model, the modification term seems to have no straightforward physical motivation. Arguably one could invoke linear perturbation theory to explain the asymptotically linear behavior in the high-density regime \citep{Desjacques2018}, but $4h^{-1}$-Mpc scales, at which the modification term still provides a good fit, are well-outside the linear regime. In a subsequent paper (Repp \& Szapudi, in prep.) we explore whether or not the modification is useful in characterizing empirical galaxy data.

\subsection{Best Fits to Abundance-Matched Data}
\label{sec:AM}
The procedure in Section~\ref{sec:dircomp} demonstrates that, for this simulation, the Ising model provides a superior description (compared to standard bias models) of the small-scale galaxy distribution. In addition, the Ising model (without the modification term of Section~\ref{sec:modIsing}) requires no more free parameters than the quadratic model.

However, adapting this procedure to empirical galaxy counts is problematic since dark matter densities are typically unavailable. We thus consider an alternative test relying on our understanding of the dark matter probability distribution $\mathcal{P}(A)$. Here, as in the previous sections, we use the distribution $\mathcal{P}(A)$ derived from the Millennium Simulation results; in dealing with observational data, one could use the log-GEV prescription of  \citet{ReppApdf}.

The alternative approach explored here is to approximate the relationship between $A = \ln(1+\delta)$ and $N_\mathrm{gal}$ using an abundance matching procedure in which we match the cumulative distribution functions of $A$ and $N_\mathrm{gal}$. Thus, we define $N_\mathrm{AM}(A)$ as the lowest value which satisfies the inequality
\begin{equation}
\mathcal{F}(A) \equiv \int_{-\infty}^A dA'\,\mathcal{P}(A') \le \sum_{N_\mathrm{gal}=0}^{N_\mathrm{AM}(A)} \mathcal{P}(N_\mathrm{gal}) \equiv \mathcal{F}(N_\mathrm{AM}(A)).
\label{eq:AM}
\end{equation}

We then assume that for an underlying dark matter density $A$, the mean number of galaxies $\langle N_\mathrm{gal} \rangle_A$ equals $N_\mathrm{AM}(A)$; in essence this approximation ignores the scatter of $N_\mathrm{gal}$ about its mean.

To gauge the accuracy of this procedure, we obtain 15 evenly-spaced points (in $A$-space) between the 0.5th and 99.5th percentiles of the $A$-distribution. For these values of $A$, we then use Equation~\ref{eq:AM} to determine $N_\mathrm{AM}(A)$, which we then set equal to $\langle N_\mathrm{gal} \rangle_A$. These abundance-matched values appear as light magenta points in Figures~\ref{fig:MillSim_scat0}--\ref{fig:MillSim_scat2.1}, thus facilitating comparison with the true values (in red) of $\langle N_\mathrm{gal} \rangle_A$.

Two facts are evident: first, at small scales ($\la 5h^{-1}$ Mpc) and high values of $A$, the abundance-matched points are systematically higher than the direct-comparison points. This systematic offset is unsurprising, given that the distribution produced by Poisson scatter skews to the right. Second, we see that the abundance-matched points drop more sharply at low values of $A$ than do the true values. This fact is a direct consequence of discreteness in the distribution of $N_\mathrm{gal}$: namely, $N_\mathrm{AM}(A)$ vanishes for all values of $A$ such that $\int_{-\infty}^A dA'\,\mathcal{P}(A') \,\le \,\mathcal{P}(N_\mathrm{gal}=0)$. It is also for this reason that this second effect disappears at larger scales ($\ga 15h^{-1}$ Mpc) where discreteness is less pronounced.

The abundance-matched points still display sigmoid behavior. In general, the transition slope (parametrized by $T$) between the low and high asymptotic values is sharper for the abundance-matched points, precisely because of the aforementioned cutoff. A fit to these abundance-matched points, though crude, would still avoid the unphysical behaviors endemic to the linear and quadratic models at small scales.

We thus conclude that the abundance-matching procedure gives us a reasonable qualitative understanding of the relationship between $A$ and $N_\mathrm{gal}$, with the advantage that we need not actually determine the dark matter density in each survey cell. However, reliance on abundance-matching will typically exaggerate the sharpness of the transition between the low- and high-$N_\mathrm{gal}$ regimes; thus, while abundance-matching seems capable of discriminating between classes of models (e.g., Ising vs. linear) at small scales, it is nevertheless suboptimal for actual model-fitting. We present a better alternative in future work (Repp \& Szapudi, in prep.).

\section{Discussion}
\label{sec:disc}
It is first worthwhile to re-emphasize the stochastic nature of this model, in that it predicts mean values only. This aspect in fact compensates for the approximate nature of the assumptions outlined at the start of Section~\ref{sec:model}. The Poisson distribution is perhaps the most natural assumption for the distribution of $N_\mathrm{gal}$ given $A$; however, the Ising model is compatible with other prescriptions for $\mathcal{P}(N_\mathrm{gal}|A)$ as well.

\begin{figure}
\leavevmode\epsfxsize=8cm\epsfbox{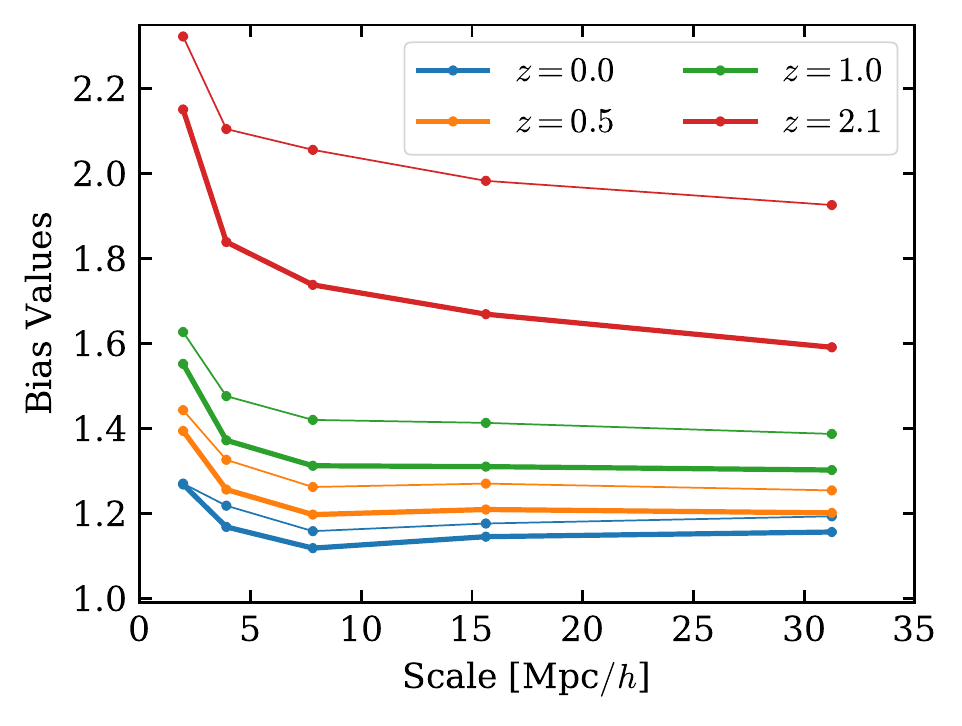}
\caption{Best-fitting Ising bias values (as a function of scale); thick and thin lines show results for the plain and modified Ising models, respectively.}
\label{fig:bias_vals}
\end{figure}

Second, it seems that, at the smallest scale ($2.0h^{-1}$ Mpc) in Figures~\ref{fig:MillSim_scat0}--\ref{fig:MillSim_scat2.1}, we observe a ``bump'' in $M$ just to the right of the sharp rise (although the bump may not be present for $z=2.07$). The feature seems to be a real effect not captured by any of the models here considered. It is tempting to speculate that there exists an ``optimal'' density for galaxy formation, above which the early formation of galaxies in multiple subhaloes suppresses subsequent formation in neighboring sites. (In constrast, we assumed that galaxy formation is strictly local.) It is also possible that this effect is peculiar to the particular semi-analytic model behind our galaxy catalog; it is even possible that the effect might depend on the stellar mass cutoff employed in selecting galaxies for the catalog.

Next, it is instructive to consider the trends in the best-fitting Ising bias parameters ($b$ in Equations~\ref{eq:final_FD} and \ref{eq:mod_Ising}), shown in Figure~\ref{fig:bias_vals}. In almost every case, the bias values from the modified model are higher than those from the original model; nevertheless, they show similar trends. First, the bias increases monotonically with redshift (as expected, given that the common motion of both dark matter and galaxies should reduce the bias over time, \citealp{TegmarkPeebles1998}). Second, at scales larger than $5h^{-1}$ Mpc, the best-fitting values vary with scale by only a few per cent (with the exception of $z = 2.1$). A similar behavior appears in fig. 2 of \citet{Contarini2019}, where the values of $\delta$ and $\delta_g$ from different scales lie on the same curve. \citet{Contarini2019} consider scales larger than those in this work; it appears that at much smaller scales, the bias is \emph{not} scale-independent but rises sharply.

\begin{figure}
\leavevmode\epsfxsize=8cm\epsfbox{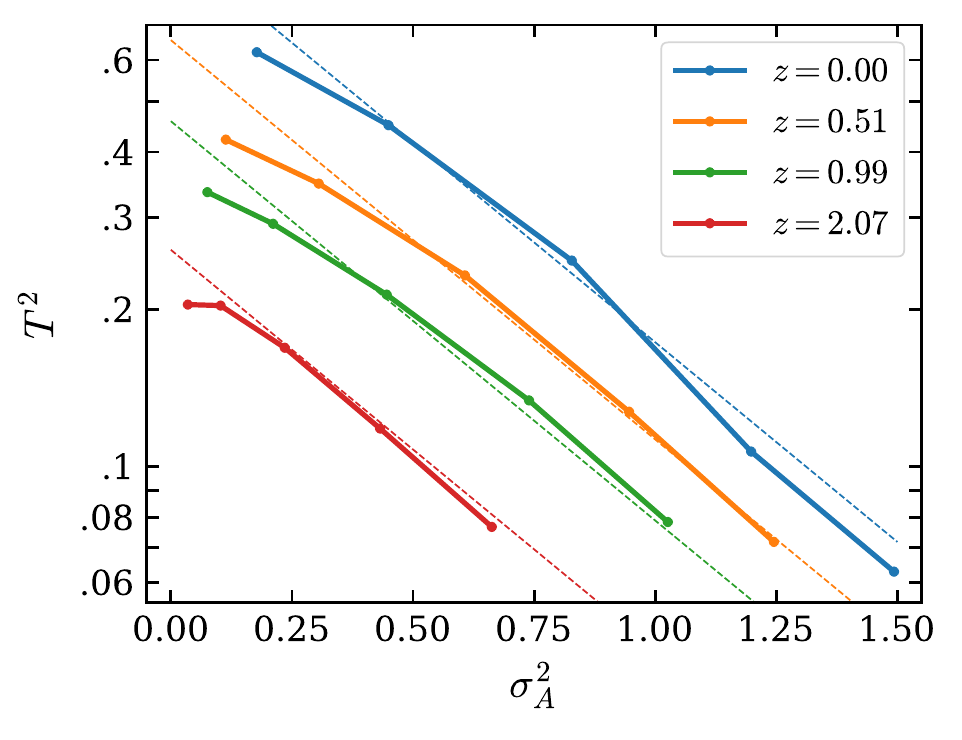}
\caption{Dependence of the best-fitting value of $T$ (the pseudo-temperature Ising parameter) on the variance of log density. The dashed lines show an exponential approximation at each redshift, with the offsets calculated from Equation~\ref{eq:GF}.}
\label{fig:T_trend}
\end{figure}

Finally, let us consider the physical origin of the pseudo-temperature parameter $T$. In a true Fermi-Dirac situation, this value would parametrize the ability of (say) electrons to scatter from one state to another. In our case, $T$ parametrizes the various possibilities for the distribution of mass within a survey pixel: recall from Section~\ref{sec:model} that we assume greater energetic favorability for clustered subhaloes. Since a cell of log density $A\equiv \ln(1+\delta)$ can host a variety of matter configurations (number of subhaloes, degree of clustering, etc.), a given value of $A$ corresponds to a range of energetic favorabilities. Just as $T$ (or, strictly speaking, $-T$) is analogous to temperature, the range of matter distributions internal to a given cell is analogous to entropy. In other words, $T$ parametrizes the thermodynamic effect of coarse-graining. 

We show in Appendix~\ref{app:Trel} that, for a given redshift, $T^2$ should be a decreasing function of the log variance $\sigma_A^2 = \langle A^2 \rangle - \langle A \rangle^2$. Figure~\ref{fig:T_trend} displays the actual best-fitting values of $T$ (from the unmodified Ising model), and we see that this expectation is justified. At large scales (low variances) $T$ rises, producing the shallower transitions seen in the lower panels of Figures~\ref{fig:MillSim_scat0}--\ref{fig:MillSim_scat2.1}. Physically, this behavior reflects that fact that larger cells accommodate a wider range of internal matter configurations. At small scales (high variances) $T$ approaches zero, producing the sharper transitions in the earlier panels of those figures. Again, the physical interpretation of this trend is that the survey cell size is closer to the typical subhalo scale. In general, of course, the value of $T$ depends on both the type of galaxy under consideration and the specific survey parameters.

It is also apparent in Figure~\ref{fig:T_trend} that the relationship between $T^2$ and $\sigma_A^2$ is approximately exponential, and that a change in redshift simply offsets the relationship by a multiplicative factor (to first order) without significantly changing the shape. It turns out for a given $\sigma_A^2$, the height of the curve varies linearly with the growth function $D^2(z)$. For instance, taking as fiducial the values at $\sigma_A^2 = 0.50$ (since that point is in the domain of all four plotted curves), we find that
\begin{equation}
T_z^2(\sigma_A^2=0.5) = 0.378D^2(z) + 0.039,
\label{eq:GF}
\end{equation}
where $D^2(z)$ is the amplitude of the linear power spectrum normalized such that $D^2(0) = 1$. If we approximate the $\sigma_A^2$-$T^2$ relationship (at each redshift) by an exponential function and scale using Equation~\ref{eq:GF}, then we obtain the dashed lines in Figure~\ref{fig:T_trend}. In Appendix~\ref{app:Trel} we provide theoretical justification for both the exponential dependence of $T^2$ on $\sigma_A^2$ and for the form of the $z$-dependence in Equation~\ref{eq:GF}.

\section{Conclusions}
\label{sec:concl}
A reliable model of galaxy bias is necessary in order to interpret the data from dense galaxy surveys. Conventional bias models (such as the linear or quadratic) work well in high-density regions, but they yield unphysical results for voids, which contain significant cosmological information.

We have here presented an Ising model that avoids unphysical predictions. This model follows from a small set of simple physical assumptions about galaxy formation. We have tested the model using Millennium Simulation galaxy catalogs and have found it vastly superior to conventional models on scales approximately $2$--$10h^{-1}$Mpc, although the linear and quadratic models are preferable at scales above $30h^{-1}$Mpc. (In the intermediate regime we find that none of the models seems to describe the bias well, although the Ising model is still typically preferable to the others.) At all scales considered, however, the Ising model provides reasonable results in both low- and high-density regions and furthermore requires only two free parameters (as does the quadratic model).

At high densities, the Ising model yields the correct zeroth-order (constant) asymptotic behavior. We also considered (in Section~\ref{sec:modIsing}) the possibility of modifying the Ising model to guarantee the same first-order asymptotics as the linear model. The additional term produces a lower reduced $\chi^2$ at intermediate scales but does not seem to improve the fit at the smallest ($2h^{-1}$ Mpc) and largest ($30h^{-1}$ Mpc) scales. This behavior -- as well as a possible physical motivation for the modification -- merits further investigation.

In this work we have restricted our tests of the Ising model to simulation data alone. We are currently (Repp \& Szapudi, in prep.) testing the model against empirical galaxy survey data. Nevertheless, the results presented here already seem to demonstrate the superiority of the Ising model for the analysis of galaxy surveys at non-linear scales. 

\section*{Acknowledgements}
The Millennium Simulation data bases used in this work and the web application providing online access to them were constructed as part of the activities of the German Astrophysical Virtual Observatory (GAVO). This work was supported by NASA Headquarters under the NASA Earth and Space Science Fellowship program -- ``Grant 80NSSC18K1081'' -- and AR gratefully acknowledges the support. IS acknowledges support from National Science Foundation (NSF) award 1616974.

\bibliographystyle{astron}
\bibliography{Thesis_Proposal}

\appendix
\section{Determining Error on $M$}
\label{app:err}
The vertical axes of Figures~\ref{fig:MillSim_scat0}--\ref{fig:MillSim_scat2.1} display the values of $M(A) = \langle N_\mathrm{gal} \rangle_A \cdot e^{-A}$. (For the remainder of this appendix, we write $N$ for $N_\mathrm{gal}$.)

In a bin $\mathcal{A}$ of values of $A$, it is straightforward to show that the average value of $M$ in the bin is
\begin{equation}
\langle M \rangle_\mathcal{A} = \sum_N \int_\mathcal{A} dA \,\, \mathcal{P}_\mathcal{A}(N,A) \, Ne^{-A},
\end{equation}
where $\mathcal{P}_\mathcal{A}(N,A)$ denotes the joint probability distribution within the bin $\mathcal{A}$. Hence, given
a set of measurements $N_i$, $A_i$ within an $A$-bin, the average $\hat{M} \equiv \langle N_i e^{-A_i} \rangle$ is the proper estimator for $\langle M \rangle_\mathcal{A}$, the expected value of $M$ within the bin $\mathcal{A}$; likewise, $\langle N_i^2 e^{-2A_i} \rangle$ correctly estimates $\langle M^2 \rangle_\mathcal{A}$. Thus the variance $\sigma_{M_i}^2$ of $N_i e^{-A_i}$ properly estimates $\sigma_M^2$; and thus the uncertainty of our estimator $\hat{M}$ is $\sigma_{\hat{M}} = \sigma_{M_i}/\sqrt{N_\mathrm{cells}}$, where $N_\mathrm{cells}$ is the number of survey cells in the bin $\mathcal{A}$. $\hat{M}$ and $\sigma_{\hat{M}}$ provide the vertical positions and vertical error bars in Figures~\ref{fig:MillSim_scat0}--\ref{fig:MillSim_scat2.1}.

The preceding analysis assumes that the measurements $N_i$, $A_i$ within a bin $\mathcal{A}$ are independent and thus uncorrelated; however, significant correlation does exist between cells on small scales. Proper accounting for this correlation would require a model of the two-point probability distribution as well as a model of galaxy bias -- leading to circularity, since a model of galaxy bias is precisely what we attempt to validate in this work.

Thus, in this work we use only a subset of the dark matter and galaxy catalogs, chosen such that all cells are at least $\sim 10h^{-1}$ Mpc distant from each other; in this way we seek to minimize the correlations among cells and thus to ensure that they do not significantly influence the size of $\sigma_{\hat{M}}$. In particular, from the $1.95h^{-1}$-Mpc catalogs, we use only every fifth cell (in each dimension); from the $3.91h^{-1}$-Mpc catalogs, every third cell; and from the $7.81h^{-1}$-Mpc catalogs, every other cell. For the larger-scale catalogs, the cell centers are already at least $10h^{-1}$ Mpc apart, and so we use the full catalogs.

\section{The Relationship of $T$ and $\sigma_A$}
\label{app:Trel}
The Ising-model parameter $T$ (analogous to negative temperature) is the result of coarse-graining (see Section~\ref{sec:disc}): for a survey cell of (log) density $A$, there exist multiple possible internal matter configurations yielding that density, and each matter configuration could potentially result in a different number of galaxies. Thus, there is an inherent scatter in the relationship $A \longmapsto N$, and $T$ parametrizes this scatter. This rationale for the role of $T$ suggests that we consider the relationship between fine-grain and coarse-grain variability.

\begin{figure}
\leavevmode\epsfxsize=6cm\epsfbox{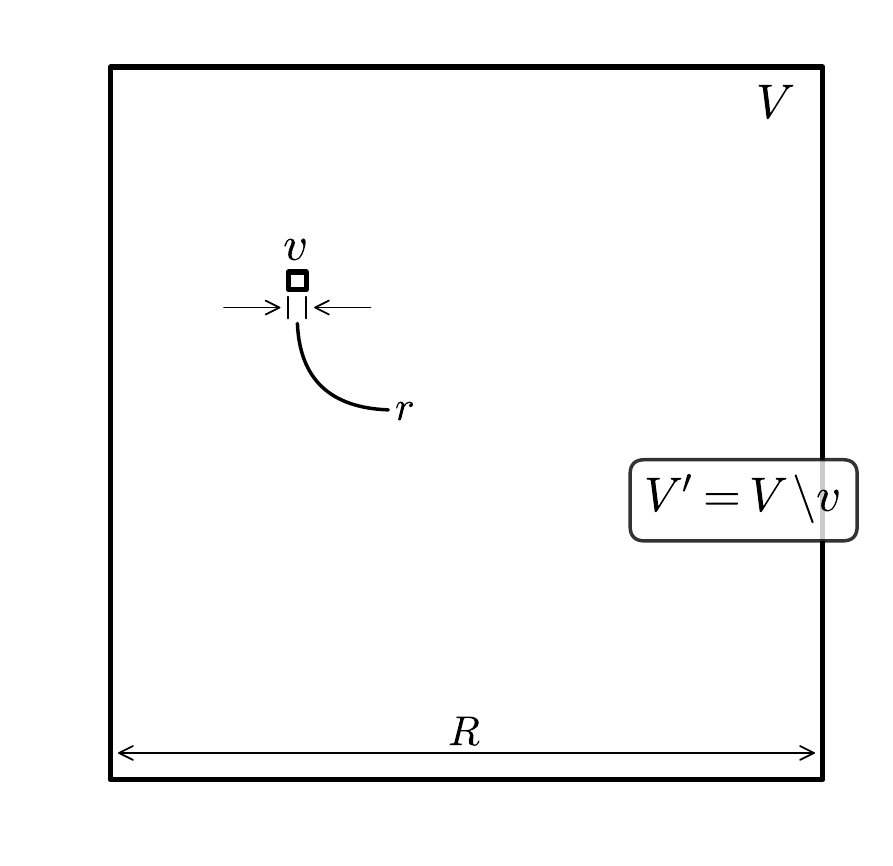}
\caption{Coarse- and fine-grain cells for Appendix~\ref{app:Trel}.}
\label{fig:boxfig}
\end{figure}

Therefore, as in Figure~\ref{fig:boxfig}, consider a survey cell $V$ of side length $R$. Subdivide $V$ into many smaller cells of side length $r$, where $r$ approximates the scale of an individual subhalo. Choose a specific small cell $v$, and let $V'$ be the complement of $v$ in $V$ (i.e., $V' = V \setminus v$).

Let $\delta_R$ be the density of $V$ (i.e., smoothed with length $R$), and $\delta_r$ be the density of $v$ (smoothed with length $r$). Then the density in the region $V'$ is
\begin{equation}
\delta_{R'} = \frac{V\delta_R - v\delta_r}{V-v},
\end{equation}
so that
\begin{align}
\delta_{R'}\delta_r & = \frac{V\delta_R\delta_r - v \delta_r^2}{V-v}\\
\langle \delta_{R} \delta_r \rangle & = \frac{V-v}{V} \langle \delta_{R'} \delta_r \rangle + \frac{v}{V} \langle \delta_r^2 \rangle,\label{eq:dRr}
\end{align}
where the average is taken over the entire survey volume (i.e., all cells $V$ containing cells $v$). We also have
\begin{align}
\delta_{R'}\delta_r & = \frac{1}{(V-v)v} \int_{V'} d^3r' \, \delta(r') \int_v d^3r\, \delta(r)\\
\langle \delta_{R'}\delta_r \rangle & = \frac{1}{(V-v)v} \int_{V'} d^3r' \int_v d^3r \, \langle \delta(r')\delta(r) \rangle \\
  & = \frac{1}{(V-v)v} \int_{V'} d^3r' \int_v d^3r \, \xi(r'-r),\label{eq:intformn}
\end{align}
where $\xi(r)$ is the two-point correlation function. Making some approximations in the limit of small $v$, Equation~\ref{eq:intformn} implies 
\begin{align}
\langle \delta_{R'}\delta_r \rangle & \approx \frac{1}{V-v} \int_{V'} d^3r' \, \xi(r') \\
  & \approx \frac{V}{V-v}\sigma_R^2 - \frac{v}{V-v} \sigma_r^2.\label{eq:dRpr}
\end{align}
It follows from Equations~\ref{eq:dRr} and \ref{eq:dRpr} that the correlation $\rho_{Rr}$ between large- and small-scale density is simply 
\begin{equation}
\rho_{Rr} = \frac{\xi_{Rr}}{\sigma_R\sigma_r} = \frac{\langle \delta_R \delta_r \rangle}{\sigma_R\sigma_r} \approx \frac{\sigma_R^2}{\sigma_R\sigma_r} = \frac{\sigma_R}{\sigma_r}.
\label{eq:cor_Rr}
\end{equation}
(Note that the implication $\xi_{Rr} = \sigma_R^2$ is as expected in the limit of small $v$.) As a result of \ref{eq:cor_Rr}, the fine-grain variability within a given coarse-grain cell is
\begin{equation}
\left. \sigma_r^2 \right|_R = \sigma_r^2 \left(1 - \rho_{Rr}^2 \right) \approx \sigma_r^2 - \sigma_R^2.
\label{eq:sigrR}
\end{equation}
The quantity $\left. \sigma_r^2 \right|_R$ is the variance of the fine-grained matter density within a coarse-grained cell, and we thus identify it with $T^2$ and note that it is a decreasing function of $\sigma_R^2$.

Now, our first assumption in constructing the Ising model is that galaxy formation is determined (to first order) by the initial conditions; thus, for the variance $\sigma_R^2$ we should, strictly speaking, use the initial variance or, as evolved forward in time, the linear variance $\sigma^2_\mathrm{lin}$. However, the log variance $\sigma_A^2$ is an increasing function of $\sigma^2_\mathrm{lin}$ \citep{Repp2017}, and thus we can conclude that $T^2$ is a decreasing function of $\sigma_A^2$.

Indeed, we can go further to explain the approximately exponential relationship observed in Figure~\ref{fig:T_trend}, in which $\ln T^2$ is roughly a linear function of $\sigma_A^2$. The key to doing so is the relationship between $\sigma_A^2$ and the linear variance, $\sigma_\mathrm{lin}^2$, developed in \citet{Repp2017}, where we show that
\begin{equation}
\sigma_\mathrm{lin}^2 = \mu \exp \frac{\sigma_A^2}{\mu} - \mu,
\end{equation}
with $\mu = 0.73$. So, identifying $\left. \sigma_r^2 \right|_R$ in Equation~\ref{eq:sigrR} as $T^2$ and explicitly specifying the linear variance, we can write
\begin{equation}
T^2 = \sigma_{\mathrm{lin},r}^2 - \sigma_{\mathrm{lin},R}^2,
\end{equation}
where again $r$ and $R$ specify the fine- and coarse-graining scales, respectively. Since we are measuring $\sigma_A^2$ at the coarse-grain scale, we now write
\begin{equation}
T^2 = \left(\sigma_{\mathrm{lin},r}^2 + \mu\right) - \mu \exp \frac{\sigma_A^2}{\mu},
\label{eq:expsA2}
\end{equation}
so that $T^2$ has an exponential dependence on $\sigma_A^2$, as seen in the figure. (The first term of Equation~\ref{eq:expsA2} introduces the slight non-linearity observed in the curves of Figure~\ref{fig:T_trend}.)

Next, let us examine the effect of changing the redshift. We note that the linear variance scales by $D^2(z)$, where $D(z)$ is the growth function normalized to unity at $z=0$. So if $\sigma_{r,0}$ is the linear variance at the fine-grain scale at $z=0$, we have from Equation~\ref{eq:expsA2}
\begin{equation}
T^2_z(\sigma_A^2) = \left( D^2(z)\cdot \sigma_{r,0}^2 + \mu \right) - \mu \exp \frac{\sigma_A^2}{\mu},
\label{eq:T2z}
\end{equation}
which is linear in $D^2(z)$, as in Equation~\ref{eq:GF}.

Finally, Figure~\ref{fig:T_trend} indicates that the effect of changing the redshift is, to first approximation, a simple multiplicative scaling independent of $\sigma_A^2$. If we let $c(z) \equiv D^2(z)\cdot \sigma_{r,0}^2 + \mu$, we can from Equation~\ref{eq:T2z} write
\begin{align}
\ln T^2_z(\sigma_A^2) & = \ln c(z) + \ln \left(1 - \frac{\mu}{c(z)} \exp \frac{\sigma_A^2}{\mu} \right)\\
        & \approx \ln c(z) - \frac{\mu}{c(z)} \exp \frac{\sigma_A^2}{\mu},
\end{align}
since we take the fine-graining scale to be small enough that $\sigma^2_{r,0}$ is quite large, so that $c(z) \gg \mu = 0.73$. Then
\begin{equation}
\ln \frac{T_{z_1}^2}{T_{z_2}^2} = \ln \frac{c(z_1)}{c(z_2)} + \mu \left( \frac{1}{c(z_2)} - \frac{1}{c(z_1)} \right) \exp \frac{\sigma_A^2}{\mu}.
\label{eq:T_rat}
\end{equation}
However, the large size of $\sigma^2_{r,0}$, to which we already appealed, means that $1/c(z)$ will be small, and $1/c(z_2) - 1/c(z_1)$ will be smaller yet. It follows that the dependence on $\sigma_A^2$ in \ref{eq:T_rat} is quite weak, so that, to a good approximation, the ratio $T_{z_1}^2/T_{z_2}^2$ depends only on the redshifts involved and not on $\sigma_A^2$, which is precisely what we see in Figure~\ref{fig:T_trend}. In particular, it is the effect of redshift evolution on the fine-grain scale, as reflected in $c(z)$, which shifts the curves in Figure~\ref{fig:T_trend} down at higher redshifts.
\label{lastpage}
\end{document}